\documentclass[aps,pra,twocolumn,superscriptaddress]{revtex4}

\usepackage[dvips]{graphics}
\usepackage{epsfig}
\usepackage{amsfonts}

\begin{document}
\newcommand{\beq}{\begin{equation}}
\newcommand{\eeq}{\end{equation}}
\title{Quantum Zeno effect by indirect measurement: The effect of the detector}
\author{M. G. Makris}
\affiliation{Department of Physics, University of Crete, P.O Box 2208, Heraklion 71003, Crete, Greece}
\email[e-mail:]{makris@physics.uoc.gr}
\author{P. Lambropoulos} 
\affiliation{Department of Physics, University of Crete, P.O Box 2208, Heraklion 71003, Crete, Greece}
\affiliation{ Institute of Electronic Structure and Laser, F.O.R.TH, P.O. Box 1527, Heraklion 711 10, Crete, Greece}

\date{\today}
\pacs{03.65.Xp,03.65.Yz,06.20.Dk}

\begin{abstract}  
   We study the quantum Zeno effect in the case of indirect measurement, where the detector does not interact directly with the  unstable system. Expanding on the model  of Koshino and  Shimizu  [Phys. Rev. Lett., {\bf 92}, 030401, (2004)] we consider a realistic Hamiltonian for the detector with a finite bandwidth. We also take explicitly into account the position, the dimensions and the uncertainty in the measurement of the detector. Our results show that the quantum Zeno effect is not expected to occur, except for the unphysical case where the detector and the unstable system overlap.
\end{abstract}  

\maketitle



\section{Introduction}
In a recent paper \cite{Shimizu}, Koshino and Shimizu (KS) considered the quantum Zeno effect (QZE) \cite{khalfin,Misra,wineland,kurizki}, for an exactly exponentially decaying system.  They conclude that the possibility for observing the QZE exists even in this case, where the initial deviation from exponential behaviour, thought to be of vital importance for the QZE, is absent.

 As an example, they considered a two-level atom (TLA) decaying to its ground
state by emitting a photon counted by a detector.  Through a continuous
indirect measurement of the emitted photon, they obtain the QZE even in the
extreme case where the "jump time" is zero, which leads them to the
conclusion that the QZE is easier than expected so far to occur.

Since this contrasts with conventional wisdom, we undertook a careful reexamination of the problem.  We find that it is essential to reformulate the Hamiltonian so as to account for  the influence of the finite extent of the detector, including its distance from the TLA.  Our calculations, based on a discretization technique and the numerical solution of the resulting system of differential equations, show that the QZE does not occur, except for the unphysical situation, inherent in the model of Ref.\cite{Shimizu},  in which the TLA and the detector overlap; i.e. the  detector  contains the TLA.

\section{Hamiltonian Construction}

    The system we consider follows  as close as possible the lines of Ref. \cite{Shimizu} (the same system and formalism has been   employed by KS earlier in Ref. \cite{ShimizuOld}). The unstable system,  a two-level atom (TLA) with $|g\rangle$ the ground and $|e\rangle$ the excited state, is initially in  $|e\rangle$  and decays to its ground state by emitting a photon. The emitted photon is subsequently detected and the ``observer'' becomes aware of the TLA decay. The total quantum system we consider includes, besides the TLA and the electromagnetic field, a part of the measuring apparatus, which is treated quantum mechanically. 

  The system Hamiltonian ($\hbar=c=1$) in the form employed by KS is: 
\beq
{\mathcal H}={\mathcal H_0}+ {\mathcal H_1}+ {\mathcal H_2}
\eeq
\begin{eqnarray}
{\mathcal H_0}&=&\Omega|e\rangle\langle e|,\\
{\mathcal H_1}&=&\int d{\bf k}\left[ (\xi_{\bf k} |e\rangle\langle g|b_{\bf k} + H.c.) + k b_{\bf k}^\dagger b_{\bf k}  \right], \\
{\mathcal H_2}&=&\!\!\! \int\!\!\!\!\int d{\bf k}d\omega \left [ (\sqrt{\eta_k}  b_{\bf k}^{\dagger} c_{{\bf k} \omega} + H.c.) + \omega c^{\dagger}_{{\bf k} \omega} c_{{\bf k} \omega} \right ]
\end{eqnarray}
where ${\mathcal H_0}$ is the part representing the free evolution of the TLA, ${\mathcal H_1}$  the atom-photon interaction and the free evolution of the photon, with $b_{\bf k}$  the annihilation operator for the photon of $\bf k$ wave vector. The combined system ${\mathcal H_0}+ {\mathcal H_1}$ is coupled to a (macroscopic) detector, a part of which is modeled by ${\mathcal H_2}$ which represents quantum mechanically the measuring procedure, i.e. the detection of the emitted photon. $\xi_k$ and $\sqrt{\eta_k}$ are  the atom-photon and the  photon-detector couplings, respectively. All photon modes are coupled with the continuum of the bosonic elementary excitations in the detector, with annihilation operator  $c_{{\bf k} \omega}$. The usual commutation relations for the $b,c$ operators hold.
\begin{figure}[!h]
\centerline{\hbox{\psfig{figure=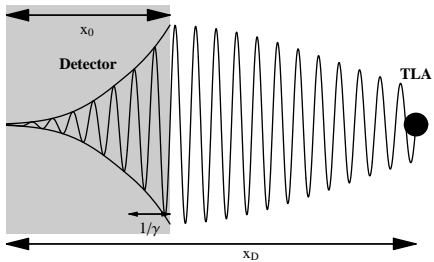,width=6cm}}}
\caption
{
A schematic plot of the TLA  and the detector (shaded region) as we consider it. The electromagnetic field emitted by the TLA enters the detector with a penetration depth of $1/\gamma$. The  atom  is displaced by $x_D$ and the detector spans from $0$ to $x_0$.
} 
\label{fig:DetectorDiag}
\end{figure}

   We wish to elaborate on two issues on this form of the Hamiltonian. First, in ${\mathcal H_2}$, the detection process is accomplished by  transferring  a quantum of a photon mode  to the detector modes through the  term ($ b_{\bf k}^{\dagger} c_{{\bf k} \omega}$) which conserves  $\bf k$. This means that there is no  uncertainty  in the detection process of  $\bf k$ which is a rather unphysical feature.  Consider a detector capable of detecting  (practically) all photons. In the case that the electromagnetic field  decays inside the detector as  $e^{-\gamma x}$ (see Fig. \ref{fig:DetectorDiag}), the momentum of the detected photon is determined within $\gamma$ (from the uncertainty relations $\Delta x \Delta k \sim 1$). Thus, there is an inherent uncertainty on the outcome of the measurement, a photon with $k$ wavenumber can be detected as $k^{\prime}$ inside the bandwidth $\gamma$. We take this into account  by introducing $\mathcal{C}(k,k^{\prime})$ in  ${\mathcal H_2}$  which becomes:
\begin{eqnarray}
{\mathcal H_2}&=&\int\!\!\!\!\int\!\!\!\!\int d{\bf k^{\prime}}d{\bf k}d\omega \left(\mathcal{C} (k,k^{\prime})   b_{\bf k}^{\dagger} c_{{\bf k^{\prime}} \omega} + H.c.\right) + 
\int\!\!\!\!\int d{\bf k}d\omega \phantom{2} \omega c^{\dagger}_{{\bf k}\omega} c_{{\bf k} \omega}.
\end{eqnarray}
and $\mathcal{C}$ depends on the details of the electromagnetic field attenuation inside the detector. 

  The derivation of $\mathcal{C}$ can be accomplished via two different pathways. In the first, we consider the  macroscopic characteristics of the decay of the electromagnetic field inside the detector and obtain  $\mathcal{C}$ phenomenologically. In  the second, we need to specify the details of the detector and we can derive  $\mathcal{C}$ through this more rigorous approach. Both ways lead to the same result for  $\mathcal{C}$ (physically), i.e. a smooth function with finite width, which is actually the only important factor in our model. We  briefly describe both.

   In the phenomenological approach  we can assume that the electromagnetic field attenuation inside the detector depends on two factors, the coupling strength ($\sqrt{\eta_k}$) and the  density $\rho(x)$ of the bosonic excitations of the detector. The latter is introduced to account for a smooth transition at the surface to the bulk density and/or for other space dependent particulars of the detector. The local attenuation rate of the mode of the  electromagnetic field is proportional to $\sqrt{\eta_k  \rho(x)}$ and the electromagnetic field   mode (in one dimension) becomes:
\begin{displaymath}
\mathcal{P}_k(x) = \left\{ \begin{array}{ll} 
\mathcal{N} e^{- i k x } & \textrm{, $ x >  x_0 $  } \\
\mathcal{N} e^{- i k x }e^{-\int_{x}^{x_0} \sqrt{\eta_k  \rho(x)} dx } & \textrm{, $ x \le x_0 $  } \\
\end{array} \right. ,
\end{displaymath}
where $\mathcal{N}$ is the normalization factor for the photon eigenmode, and $x_0$ is the point where the detector begins (see Fig \ref{fig:DetectorDiag}).

 The coupling of the electromagnetic field modes $\phi_k$ whith the detector modes has to be such that their decay inside the detector is of the form of $\mathcal{P}_k$. A similar approach is followed in Ref.\cite{Makris:2004} in the context of absorbing boundaries in spectral methods, where it is shown that the couplings $ \mathcal{C} (k,k^{\prime})$, there the coefficients of the absorbing boundary linear transform, are the projection coefficients of  $\tilde\mathcal{P}_k$ on  $\phi_k$, where  $\tilde\mathcal{P}_k = \mathcal{P}_k - \phi_k$, i.e. the part of the  $\mathcal{P}_k$ mode transferred to the detector. In the following numerical analyses, we assume a Gaussian attenuation inside the detector, which leads in a Gaussian $ \mathcal{C} (k,k^{\prime})=\mathcal{C} (k-k^{\prime})$.

 In the case one wishes to take into acount all the details of the detector in a more fundamental level the Hamiltonian of the detector and the resulting eigenmodes have to be specified. Then the coupling whith the  $\phi_k$  modes is $\langle \phi_{k^{\prime}}  | \mathcal{D}  |  \Phi_{k\omega} \rangle$
where $\Phi_{k\omega}$  are the eigenmodes of the detector and  $\mathcal{D}$ the coupling  operator of the detector with the electromagnetic field. We have to bear in mind that the eigenmodes of the detector are spatially localised, i.e. inside the detector. Also, since we wish to represent a detector and not a mirror the eigenmodes have to attenuate smoothly at the surface of the detector. Clearly the exact calculation has to proceed by an explicit formulation of the detector Hamiltonian and determination of its eigenmodes. We do not intend to proceed in this direction since our scope in this paper is to demonstrate the qualitative effects of the detector width and position of the obervation of QZE. The basic result of such an analysis can be deduced  by considerind a simple form  for the $\Phi_{k\omega}$ in conformation with the two restrictions we mentioned: space localisation and smooth variations, for example a plane wave with a Gaussian envelope. In this simple case it is evident that the $\mathcal{C} (k,k^{\prime})$ could  practically be thought of as  a Gaussian.

 The ${\mathcal H_2}$ in \cite{Shimizu} is a subclass of this generalized version with $ \mathcal{C} (k,k^{\prime})$ being a delta function.  In retrospect, this means that $\gamma \to 0$, which implies that the physical dimensions of the detector tend to infinity. The latter is a direct artificial influence on the dynamics of the decaying two-level atom, since it implies that the (infinite) detector and the TLA overlap. We  return to this issue latter on.

   The second issue we wish to  take into account is the relative position of the detector and the TLA. This is straightforward, and is accomplished by including the correct displacement phase factor in the Hamiltonian. This phase has the simple form  $\phi_k = e^{i k x_D }$ (see Fig. \ref{fig:DetectorDiag}), as employed for example  in \cite{Meystre} for the somewhat similar case of a TLA coupled through the electromagnetic field with another TLA. The way the Hamiltonian is written so far, the TLA and the detector overlap  and  we have to displace one of them. It is more convenient to displace the atom, since it involves inclusion of the phase factor in fewer terms, so the term  ${\mathcal H_1}$ of the  Hamiltonian becomes:
\begin{eqnarray}
{\mathcal H_1}&=&\int d{\bf k}\left[ (\xi_{\bf k} e^{i k x_D} |e\rangle\langle g|b_{\bf k}+ H.c.) + k b_{\bf k}^\dagger b_{\bf k}  \right].
\end{eqnarray}

    In general the displacement is determined by the problem at hand, but in all cases it should be such that the atom does not overlap with the detector. In the present form of the Hamiltonian the detector is  at $x=0$.

\section{Discretisation}

   We model the  electromagnetic field  and the modes of the detector with a set of discrete modes. The wavefunction of the system can be written as:
\beq
|\psi(t)\rangle=\alpha|e,0,0\rangle + \sum_{k}b_{k}|g,1_{k},0\rangle + \sum_{k,\omega}c_{k,\omega}|g,0,1_{k,\omega}\rangle
\eeq 
where the states involved are product states and, for instance, $|g,1_k,0\rangle=|g\rangle|1_k\rangle|0\rangle$ where $|1_k\rangle$ means one photon emitted in the $k$-th mode  and $|0\rangle$ is the zero-quanta state of the detector.

   Thus, the initial state vector of the system is $|e,0,0\rangle$ and the amplitudes obey the Schr\"odinger equation:
\begin{eqnarray}
i \dot{\alpha}&=& \Omega \alpha + \sum_k e^{i k  x_D}\xi_k b_{k} \\
i \dot{b}_k&=& \omega_k b_{k} + \xi_k e^{-i k  x_D}\alpha  +  \sum_{k^{\prime},\omega} \sqrt{\eta_{k^{\prime}}}\mathcal{C}(k,k^{\prime}) c_{k^{\prime},\omega}\\
i \dot{c}_{k,\omega}&=& \omega c_{k,\omega} + 
              \sum_{k^{\prime}} \sqrt{\eta_{k^{\prime}}} \mathcal{C}(k,k^{\prime}) b_{k^{\prime}} \label{eq:c_corr}
\end{eqnarray}
where $k$ is the index of the  discrete modes used to model the electromagnetic field   and  $k,\omega$ the indexes  for the discrete modes for the {\bf k} and $\omega$ of the detector quanta. In case $k$ appears by itself, it simply is the value of  {\bf k} of the mode.

Consider for the moment the limit of our Hamiltonian that corresponds to the Hamiltonian employed in \cite{Shimizu}, i.e $\mathcal{C}(k,k^{\prime})=\delta_{k,k^{\prime}}$ and $x_D=0$. Then the differential equations for the amplitudes would be:
\begin{eqnarray}
i \dot{\alpha}&=& \Omega \alpha + \sum_{k} \xi_k b_{k} \\
i \dot{b}_{k}&=& \omega_k b_{k} + \xi_k\alpha  +  \sum_{\omega} \sqrt{\eta_{k}} c_{k,\omega}\\
i \dot{c}_{k,\omega}&=& \omega c_{k,\omega} + 
               \sqrt{\eta_{k}} b_{k}.
\end{eqnarray}

  In this set of equations $c_{k\omega}$ is coupled only to {\em one} mode of the  electromagnetic field, which means that the detector interacts immediately with the emitted photon, without allowing any time delay associated with the distance it has to travel from the TLA to the detector. On the contrary, in Eq.(\ref{eq:c_corr}) the detector modes interact with a {\em superposition} which allows for spatial localization of the interaction, accounting  thus  correctly  for the time delay and the detector position.

\begin{figure}[!t]
\centerline{\hbox{\psfig{figure=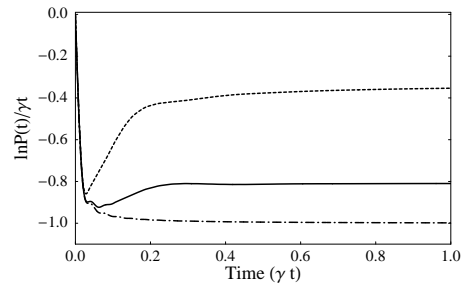,width=6cm}}}
\caption
{The decay rate of the TLA excited state population, coupled to the detector, over the free decay rate, following KS approach.  
Parameters are   $2 \pi \Delta/\gamma=100,\eta/\gamma=1,10$ for the solid and dashed curve respectively. Together we show, for the sake of comparison, the decay rate in the absence of the detector (dot-dashed curve). Since we take into account only a finite bandwidth, the decay rate is 0 at $t=0$. Discretisation range: 0 to 2$\Omega$, with  100 modes for the electromagnetic field and 100(k)$\times$100($\omega$) modes for the detector in the same range.
} 
\label{fig:Shim_Res}
\end{figure}

   We  proceed with a numerical solution for the system of differential equations. The discretisation scheme \cite{NikolopoulosG:1999,NikolopoulosG:2000}   is as simple as possible. We choose a range for {\bf $k$} and $\omega$ which we span with equidistant modes. The results are considered converged if unaltered upon increasing both the range and the density of discrete modes. Of course the choice of discretisation range is not arbitrary but based on the particulars of the problem. In this case, we take $|{\bf k}|$  around the transition frequency of the TLA and the same for $\omega$. Due to the finite interval of $|{\bf k}|$ space that we take into account, the ``jump time'' is  not infinitesimal, although it  can be made as small as computationally feasible. In any case, a non-zero $\tau$ should make QZE easier to observe.

   The situation  we have considered is equivalent to the TLA  placed at the center  of  a (hollow) spherical detector, which effectively is  a 1-D problem. In this case we have to limit to outgoing waves, thus restrict $k$  to positive values.

\section{Results}

First, we establish a direct correspondence with the results obtained in Ref.\cite{Shimizu}. We set $x_d=0$, $\mathcal{C}(k,k^{\prime})=\delta_{k,k^{\prime}}$, the atom-photon coupling  independent of ${\bf k}$ and assume that the coupling between the photon and the detector is:
\beq
\eta_k=\frac{\eta/2\pi}{1+[(k-\Omega)/\Delta]^n}
\eeq
with $\Delta$ a measure of the  photon energy  range  for which the detector is sensitive and $n$ a parameter defining the sharpness of the detector response ($n=6$). In Fig. \ref{fig:Shim_Res} we show our results which match  those obtained in Ref.\cite{Shimizu} analytically, except for a factor of 10 in the value of $\eta$, which we attribute to a possible misprint in the caption of their figure; especially since we are unable to reproduce their graph by employing their formula. The initial fast drop of the decay rate is due to the finite range of frequencies we consider in the discretization, the width of this region is of the order of $1/\Delta\Omega$, where  $\Delta\Omega$ is the bandwidth of the discretization.  After this transient region, the rise of the decay rate to its assymptotic value is resolved in accordance Figure 3 of  Ref.\cite{Shimizu}. 

\begin{figure}[!t]
\centerline{\hbox{\psfig{figure=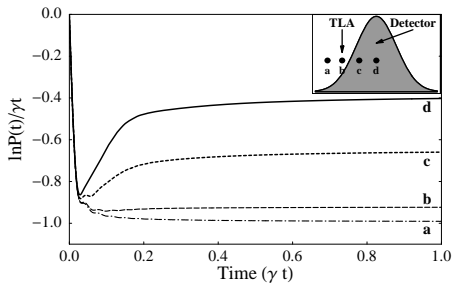,width=6cm}}}
\caption
{ The decay rate of the TLA excited state population  over the free decay rate as its  relative position with the detector is varied (inset). We considered four cases, from the TLA being at the center of the detector (a) to the TLA outside of the detector (d). The decay rate changes smoothly from the results obtained in Ref.\cite{Shimizu} to the free decay rate. The detector is assumed to have an effective  gaussian profile, with a full width half maximum of 33 ($\hbar=c=\Omega=1$), as shown in the inset. Parameters:  $2 \pi \Delta/\gamma=100,\eta/\gamma=10$, discretisation as in Fig.\ref{fig:Shim_Res}, $\mathcal{C}(k,k^{\prime})=0.103 e^{-((k-k^{\prime})/5.5)^2 }$.}
\label{fig:Pop}
\end{figure}

\begin{figure}[!t]
\centerline{\hbox{\psfig{figure=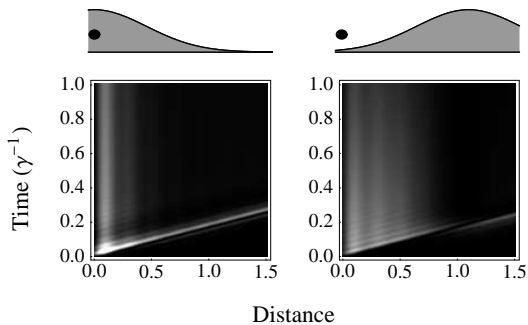,width=7.5cm}}}
\caption
{
The intensity profile of the emitted photon for the cases (a) and (d) of Fig.\ref{fig:Pop}. In the top part of each figure we show the corresponding detector profile and position. The distance is measured from the TLA and in units of detector full width half maximum. Lighter shading stands for higher inensities.
} 
\label{fig:Intensity}
\end{figure}
We proceed by considering a detector with finite width, the same in all other aspects with the detector in Ref.\cite{Shimizu}. In Fig. \ref{fig:Pop} we show the decay rate of the  population of the $|e\rangle$ state over the free decay rate, as a function of $t$ in cases where the TLA overlaps the detector and where it is  spatially  separated. When they overlap, it is evident that the decay process is decelerated, with decay rate similar to the one obtained in  Ref.\cite{Shimizu} ($\sim$0.40 vs. $\sim$0.35, case (a)). Once the TLA starts to get separated from the detector its decay rate approaches fast  the free decay rate (cases (c) and (d)). The influence of the relative position of the detector and the TLA on the dynamics of the system is shown in Fig. \ref{fig:Intensity}. The time evolution of the intensity profile  of the emitted photon shows two qualitatively distinct features. In the case where the TLA and the detector overlap, the detector captures the emitted photon instantly and acts as a ``memory''  retaining the photon close to the TLA and slowing down the decay rate. In fact, this is the effect reported in Ref.\cite{Shimizu}. Once the TLA is separated from the detector, the photon travels uninterrupted until it is absorbed by the detector. In this case the atom decays with the free space rate without being influenced by the detector.

\section{Concluding Remarks}

Starting with the formalism of Ref.\cite{Shimizu}, we modified the Hamiltonian representing the TLA,  electromagnetic field and interaction with the detector, so as to explicitly take into account the position and the spatial width of the detector.  The modified formulation allows the analysis of the realistic situation in which the detector is spatially separated from the atom, yielding the model of Ref.\cite{Shimizu} as a special case which is shown to correspond to the detector overlapping with the atom. This is actually the case of an excited atom decaying inside a dielectric \cite{Barnett:1992,Barnett:1995}.  Having calculated the decay probability of the TLA, we find that it is not affected by the measurement precedure, except in the rather unphysical situation in which the atom overlaps with the detector.  We are thus compelled to conclude that the QZE does not occur by indirect measurements, at least in the context of Ref.\cite{Shimizu}.

\acknowledgments
 The work of MM  was supported by the  HPAK$\Lambda$EITO$\Sigma$ program of E$\Pi$EAEK II, funded by the Greek Ministry of Education and the European Union.

\bibliographystyle{apsrev}

\begin{thebibliography}{10}
\expandafter\ifx\csname natexlab\endcsname\relax\def\natexlab#1{#1}\fi
\expandafter\ifx\csname bibnamefont\endcsname\relax
  \def\bibnamefont#1{#1}\fi
\expandafter\ifx\csname bibfnamefont\endcsname\relax
  \def\bibfnamefont#1{#1}\fi
\expandafter\ifx\csname citenamefont\endcsname\relax
  \def\citenamefont#1{#1}\fi
\expandafter\ifx\csname url\endcsname\relax
  \def\url#1{\texttt{#1}}\fi
\expandafter\ifx\csname urlprefix\endcsname\relax\def\urlprefix{URL }\fi
\providecommand{\bibinfo}[2]{#2}
\providecommand{\eprint}[2][]{\url{#2}}

\bibitem[{\citenamefont{Koshino and Shimizu}(2004)}]{Shimizu}
\bibinfo{author}{\bibfnamefont{K.}~\bibnamefont{Koshino}} \bibnamefont{and}
  \bibinfo{author}{\bibfnamefont{A.}~\bibnamefont{Shimizu}},
  \bibinfo{journal}{Phys. Rev. Lett.} \textbf{\bibinfo{volume}{92}},
  \bibinfo{pages}{030401} (\bibinfo{year}{2004}).

\bibitem[{\citenamefont{Khalfin}(1968)}]{khalfin}
\bibinfo{author}{\bibfnamefont{L.}~\bibnamefont{Khalfin}},
  \bibinfo{journal}{JETP Lett.} \textbf{\bibinfo{volume}{8}},
  \bibinfo{pages}{65} (\bibinfo{year}{1968}).

\bibitem[{\citenamefont{Misra and G.Sudarshan}(1977)}]{Misra}
\bibinfo{author}{\bibfnamefont{B.}~\bibnamefont{Misra}} \bibnamefont{and}
  \bibinfo{author}{\bibfnamefont{E.~C. G.} \bibnamefont{Sudarshan}},
  \bibinfo{journal}{J. Math. Phys. (N.Y.)} \textbf{\bibinfo{volume}{18}},
  \bibinfo{pages}{756} (\bibinfo{year}{1977}).

\bibitem[{\citenamefont{Itano et~al.}(1990)\citenamefont{Itano, Heinzen,
  Bollinger, and Wineland}}]{wineland}
\bibinfo{author}{\bibfnamefont{W.~M.} \bibnamefont{Itano}},
  \bibinfo{author}{\bibfnamefont{D.~J.} \bibnamefont{Heinzen}},
  \bibinfo{author}{\bibfnamefont{J.~J.} \bibnamefont{Bollinger}},
  \bibnamefont{and} \bibinfo{author}{\bibfnamefont{D.~J.}
  \bibnamefont{Wineland}}, \bibinfo{journal}{Phys. Rev. A}
  \textbf{\bibinfo{volume}{41}}, \bibinfo{pages}{2295} (\bibinfo{year}{1990}).

\bibitem[{\citenamefont{Kofman and Kurizki}(2000)}]{kurizki}
\bibinfo{author}{\bibfnamefont{A.}~\bibnamefont{Kofman}} \bibnamefont{and}
  \bibinfo{author}{\bibfnamefont{G.}~\bibnamefont{Kurizki}},
  \bibinfo{journal}{Nature} \textbf{\bibinfo{volume}{405}},
  \bibinfo{pages}{546} (\bibinfo{year}{2000}).

\bibitem[{\citenamefont{Koshino and Shimizu}(2003)}]{ShimizuOld}
\bibinfo{author}{\bibfnamefont{K.}~\bibnamefont{Koshino}} \bibnamefont{and}
  \bibinfo{author}{\bibfnamefont{A.}~\bibnamefont{Shimizu}},
  \bibinfo{journal}{Phys. Rev. A} \textbf{\bibinfo{volume}{67}},
  \bibinfo{pages}{042101} (\bibinfo{year}{2003}).

\bibitem[{\citenamefont{Goldstein and Meystre}(1997)}]{Meystre}
\bibinfo{author}{\bibfnamefont{E.}~\bibnamefont{Goldstein}} \bibnamefont{and}
  \bibinfo{author}{\bibfnamefont{P.}~\bibnamefont{Meystre}},
  \bibinfo{journal}{Phys. Rev. A} \textbf{\bibinfo{volume}{56}},
  \bibinfo{pages}{5135} (\bibinfo{year}{1997}).

\bibitem[{\citenamefont{Makris}(2000)}]{Makris:2004}
\bibinfo{author}{\bibfnamefont{M.~G.} \bibnamefont{Makris}},
  \bibinfo{journal}{Phys. Rev. E}, \textbf{\bibinfo{volume}{69}},
  \bibinfo{pages}{066702} (\bibinfo{year}{2004}).

\bibitem[{\citenamefont{Nikolopoulos et~al.}(1999)\citenamefont{Nikolopoulos,
  Bay, and Lambropoulos}}]{NikolopoulosG:1999}
\bibinfo{author}{\bibfnamefont{G.~M.} \bibnamefont{Nikolopoulos}},
  \bibinfo{author}{\bibfnamefont{S.}~\bibnamefont{Bay}}, \bibnamefont{and}
  \bibinfo{author}{\bibfnamefont{P.}~\bibnamefont{Lambropoulos}},
  \bibinfo{journal}{Phys. Rev. A} \textbf{\bibinfo{volume}{60}},
  \bibinfo{pages}{5079} (\bibinfo{year}{1999}).

\bibitem[{\citenamefont{Nikolopoulos and
  Lambropoulos}(2000)}]{NikolopoulosG:2000}
\bibinfo{author}{\bibfnamefont{G.~M.} \bibnamefont{Nikolopoulos}}
  \bibnamefont{and}
  \bibinfo{author}{\bibfnamefont{P.}~\bibnamefont{Lambropoulos}},
  \bibinfo{journal}{Phys. Rev. A} \textbf{\bibinfo{volume}{61}},
  \bibinfo{pages}{053812} (\bibinfo{year}{2000}).


\bibitem[{\citenamefont{ Huttner and
  Barnett  }(1992)}]{Barnett:1992}
\bibinfo{author}{\bibfnamefont{Bruno} \bibnamefont{Huttner}}
  \bibnamefont{and}
  \bibinfo{author}{\bibfnamefont{Stephen M.}~\bibnamefont{Barnett}},
  \bibinfo{journal}{Phys. Rev. A} \textbf{\bibinfo{volume}{46}},
  \bibinfo{pages}{4306} (\bibinfo{year}{1992}).


\bibitem[{\citenamefont{Matloob  et~al.}(1995)}]{Barnett:1995}
\bibinfo{author}{\bibfnamefont{Reza} \bibnamefont{Matloob}},
\bibinfo{author}{\bibfnamefont{Rodney} \bibnamefont{Loudon}},
\bibinfo{author}{\bibfnamefont{Stephen M.}~\bibnamefont{Barnett}}
\bibnamefont{and}
\bibinfo{author}{\bibfnamefont{John} \bibnamefont{Jeffers}},  ,
  \bibinfo{journal}{Phys. Rev. A} \textbf{\bibinfo{volume}{52}},
  \bibinfo{pages}{4824} (\bibinfo{year}{1995}).


\end{thebibliography}

\end{document}